\documentclass[floatfix,nopacs,twocolumn,amssymb,amsfonts,amsmath,aps]{revtex4}
\usepackage{graphicx,epsfig}

\usepackage{latexsym}
\usepackage{graphicx}
\usepackage{gensymb}

\begin{document}
\title{Capillary Rise of Liquids in Nanopores}
\author{Patrick Huber}
\email[Correspondence to ]{p.huber@physik.uni-saarland.de}
\affiliation{Physics and Mechatronic Engineering, Saarland University,
66041 Saarbr\"ucken, Germany}

\author{Klaus Knorr}
\affiliation{Physics and Mechatronic Engineering, Saarland University,
66041 Saarbr\"ucken, Germany}

\author{Andriy V. Kityk}
\affiliation{Institute for Computer Science, Technical University of Czestochowa, PL-42220, Poland}
\date{01/23/2006}

\begin{abstract}
We present measurements on the spontaneous imbibition (capillary rise) of water, a linear hydrocarbon (n-C$_{\rm 16}$H$_{\rm 34}$) and a liquid crystal (8OCB) into the pore space of monolithic, nanoporous Vycor glass (mean pore radius 5~nm). Measurements on the mass uptake of the porous hosts as a function of time, $m(t)$, are in good agreement with the Lucas-Washburn $\sqrt{t}$-prediction typical of imbibition of liquids into porous hosts. The relative capillary rise velocities scale as expected from the bulk fluid
parameters.
\end{abstract}
\maketitle
\section{Introduction}
Spontaneous imbibition, the capillary rise of a wetting liquid in a tube or more generally spoken a porous host, plays a crucial role in such different areas as oil recovery, printing, cooking and fluid transport in living organisms \cite{Alava2004}. It is governed by a balance of capillary action versus viscous drag and gravity forces. Therefore, the details of the imbibition process, e.g. the 
morphology and speed of the advancing imbibition front depend sensitively on 
fluid parameters (viscosities $\eta$ of displacing and displaced fluid and 
interfacial or surface tensions $\sigma$) and on the structure of the 
porous host.\\

Here, we shall present experiments on spontaneous imbibition of liquids into 
nanoporous, monolithic silica glass. We will demonstrate that such 
experiments allow one to explore flow properties of liquids in restricted 
geometry. A topic which is of importance for lubrication, transport of 
liquids through biomembranes and folding of proteins - not to mention the 
growing, technological and scientific interest concerning the physics of 
liquids in nanodevices, e.g. carbon nanotubes \cite{Urbakh2004,Drake1990,Granick2003,Gelb2002,Majumder2005,Megaridis2002,Supple2003}.\\

As porous host we have chosen ``thirsty'' Vycor glass, a nanoporous silica 
glass which is permeated by a 3D network of interconnected, randomly 
oriented pores \cite{Levitz1991}. In order to play with the complexity of the imbibing fluids we have chosen three different liquids: water, a chain-like, linear hydrocarbon (n-C$_{\rm 16}$H$_{\rm 34})$ and a rod-like liquid crystal 
(4-Cyano-4'-n-octyloxybiphenyl, 8OCB).
\begin{figure*}[!]
\epsfig{file=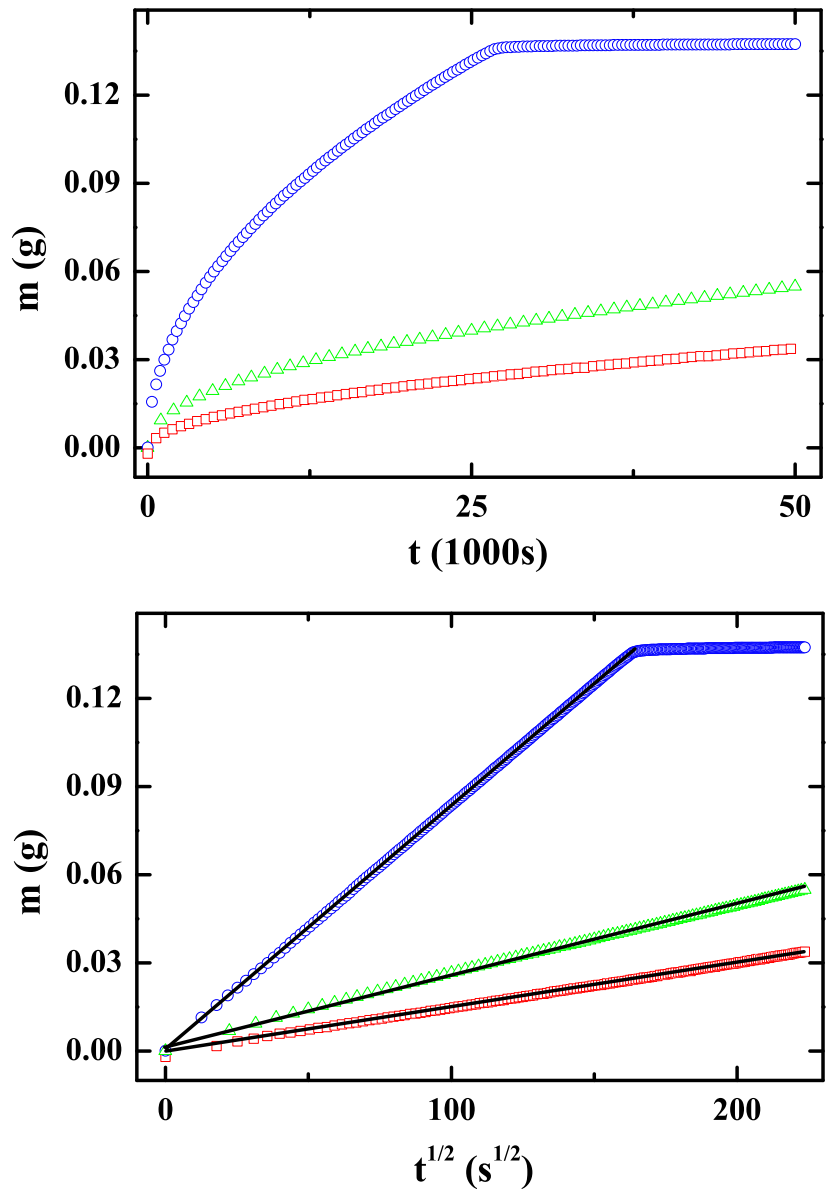, width=1.5\columnwidth}
\caption{\label{fig:1} Mass uptake of nanoporous silica (Vycor) due to the capillary rise of water (circles), n-hexadecane (triangles), and 8OCB (squares) as function of time, $t$ (upper panel), plotted versus $\sqrt{t}$ (lower panel). The solid lines in the lower panel represent fits of the data sets to the Lucas--Washburn prediction.}
\end{figure*}
\section{Experimental}
A cuboid (0.4 x 0.43 x 2.65~cm$^{\rm 3})$ was cut from a Vycor rod (Corning, code 
7930) with a diamond saw. Prior to using the Vycor sample and after each 
imbibition experiment we subjected the host to a cleaning procedure with 
hydrogen peroxide and sulphuric acid followed by rinsing in Millipore water 
and drying at 70 \r{ }C in vacuum. The cleaned glass was colorless. In 
agreement with former results on the pore structure of samples from this 
Vycor batch \cite{Wallacher2001}, the mean volume porosity, $\Phi $ and mean pore diameter, r$_{\rm m}$ was determined to $\Phi $=30{\%} and r$_{\rm m}$ =5~nm, resp.\\

Water cleaned by a Millipore filtering setup was used for the water 
imbibition experiment. N-hexadecane (99.5{\%} purity) was purchased from 
Aldrich and the 8OCB sample (99{\%} purity) from Synthon Chemicals 
(Germany). The capillary rise experiments were performed at room 
temperature, T=22\r{ }C for water and the n-alkane, whereas we have chosen a 
temperature of 85\r{ }C for the imbibition experiment on 8OCB. In the bulk 
state, the liquid crystal is at that temperature in the isotropic phase.\\ 

The mass uptake of the porous hosts as a function of time, $m(t)$, was 
recorded gravimetrically with an electronic balance after gently immersing 
the monoliths into glass vessels containing the bulk liquids. Right upon 
contact of the bottom facet of the glass matrix with the surface of the bulk 
liquid, we observed for all samples a jump in $m(t)$. It can be attributed to 
the formation of a macroscopic liquid meniscus at the bottom perimeter of 
the matrix, which happened typically within a couple of seconds after 
contact of surface and matrix. Since we are interested in the immersion of 
the liquid into the inner pore space, we will present data sets after 
subtracting this mass jump in the following. In principle, one has also to 
worry about buoyancy forces acting on the porous host and inertia effects in 
the very beginning of the imbibition process \cite{Bosanquet1923, Quere1997}. However, both contributions are negligible in our experiments, a statement supported by the data presented below.

\section{Results}
The mass uptake of the porous host by the imbibtion of water, n-C$_{\rm 16}$H$_{\rm 32}$ and 8OCB are plotted in Fig.~1 
as a function of time. For the three investigated liquids we find distinct, 
monotonic increasing mass uptake curves and the mass uptake rate, dm(t)/dt 
is monotonically decreasing with time. Due to the fast imbibition process in 
the case of water, we find additionally to the monotonic increase a 
saturation of the mass uptake at t=27100s and m$_{\rm s}$=m(27100s)=0.136g. At 
that time the advancing water front reaches the top of the porous monolith. 
The value m$_{\rm s}$ compares well with the overall mass uptake expected for a 
complete filling of the free pore space of the cuboid, 0.136g. 

The time behavior shown in Fig. 1 for the immersion of the three liquids 
into the porous monolith is well known from capillary rise experiments in 
tubes and can be derived by simple phenomenological arguments: Provided the 
liquid wets or partially wets the tube walls (contact angle, $\Theta < 
90$\r{ }), the liquid is sucked into a tube of radius r by the capillary 
pressure, p$_{\rm cap}$ , given by the Laplace equation:

\begin{eqnarray}
p_{\rm cap}= 2 \sigma  \frac{\cos{\Theta}}{r} \nonumber
\end{eqnarray}

The tube geometry calls for a Hagen-Poiseuille type flow pattern, where the 
viscous drag implies that the volume flow rate

\begin{eqnarray}
\frac{dV(t)}{dt} \sim  \frac{1}{\eta} \frac{p_{\rm cap}}{l(t)} \nonumber
\end{eqnarray}

\noindent
and thus the corresponding mass flow rate is given by

\begin{eqnarray}
\frac{dm(t)}{dt} \sim  \rho  \frac{1}{\eta}´ \frac{p_{\rm cap}}{l(t)} \nonumber
\end{eqnarray}

\noindent
, where $\rho $ is the mass density of the liquid. The variable $l(t)$ refers 
to the height of the liquid column that has already been filled up with 
liquid at time $t$. A length which is at any time proportional to $m(t)$ in a 
capillary rise experiment. Thus one arrives at a quite simple relation for 
$m(t)$ and its time derivative: 

\begin{eqnarray}
m(t) \frac{dm(t)}{dt} \sim  \rho^2 \frac{p_{\rm cap}}{\eta} \sim \rho^2   \frac{\sigma}{\eta}\cos{\Theta}
\end{eqnarray}

In a capillary rise experiment one has also to consider gravity forces, 
however, for the tiny tubes or pores investigated here, the gravity pressure 
is negligible compared to $p_{\rm cap}$. Relation 1 is solved by

\begin{eqnarray}
m(t) \sim  \rho \sqrt{\frac{\sigma}{\eta} \;\cos{\Theta}} \sqrt{t}
\end{eqnarray}

Thus the mass uptake in a capillary rise experiment should show a $\sqrt{t}$ behavior and it is sensitive to the fluid parameters ($\sigma $, $\eta $, 
$\rho$) and the fluid wall interaction (via the contact angle $\Theta$). 
Relation 2 is known as Lucas-Washburn law \cite{Lucas1918, Washburn1921} and the square of the second prefactor of relation 2 as imbibition speed, $v_{\rm i}$. Porous media, as employed in our experiments, can be considered as complex networks of simple tubes. Therefore, neglecting details of the structure of our porous 
host, such as the connectivity of the pores, their tortuosity and the pore 
size distribution, we should also expect a $\sqrt{t}$-behavior for $m(t)$. In 
agreement with this conclusion, the data points of our experiments, plotted 
versus $\sqrt{t}$, can be nicely fitted with straight lines -- see lower 
panel of Fig. 1.\\

Since we used for all three experiments the same matrix with the same pore 
microstructure and the same macroscopic extensions we can eliminate the 
unknown proportionality factors of relation 1 by a calculation of relative 
imbibition speeds, $v_{\rm ir}$. The resulting quantities should scale 
according to the ratio of the bulk fluid parameters, provided the 
continuous-type, macroscopic flow description as well as the macroscopic 
fluid parameters are valid for the capillary rise phenomenon studied. In 
order to check this, we calculated values of the bulk imbibition speeds, 
$v_{\rm i}^{\rm bulk}$ by using literature parameters for $\eta $ and $\sigma $ and contact angles on planar silica. We normalized those imbibibition speeds to the value of water:
\begin{eqnarray}
v_{\rm ir}^{\rm bulk} = v_{\rm i}^{\rm bulk} / v_{\rm i}^{\rm bulk} (\rm water) \nonumber
\end{eqnarray}
From the ratio of the square of the slopes of $m(t)$ in Fig.~1 (lower panel) divided by $\rho$, we extracted the imbibition speeds, $v_{\rm i}^{\rm conf} $, in nanoporous silica and, again, normalized them by the one of water. As one can see in Tab. 1 we get a reasonable good agreement of $v_{\rm ir}^{\rm bulk} $ and $v_{\rm ir}^{\rm conf} $ for the liquids investigated. Given the uncertainties in the literature values of the bulk fluid parameters, which we estimated to be 5-10{\%}, the small deviations between $v_{\rm ir}^{\rm bulk} $ and $v_{\rm ir}^{\rm conf} $are well within the expected error margins of $v_{\rm ir}^{\rm bulk} $.\\

It should also be noted that the decreasing mass uptake rate implicates that 
one probes also a continuous range of flow velocities in the pores during a 
given imbibition experiment, and thus a continuous range of shear rates. 
Assuming a no-slip boundary condition, one can estimate, e.g. for the water 
experiment, a decrease of the shear rate from 10$^{\rm 4}$ 1/s in the very 
beginning to 1/s at the very end of the experiment. Thus, the constant 
slopes of the Lucas-Washburn-law in Fig. 1 as a function of time implicate 
also the absence of any detectable shear-thinning or shear-thickening 
effects for the liquids studied.

\begin{table*}[htbp]
\begin{center}
\begin{tabular}{|p{55pt}|p{28pt}|p{53pt}|p{65pt}|p{38pt}|p{55pt}|p{22pt}|p{36pt}|p{46pt}|p{40pt}|}
\hline
substance& 
T \par (\r{ }C)& 
$\eta $ \par (mPas)& 
$\sigma $ \par (mN/m)& 
$\rho $ \par (g/cm$^{\rm 3}$)& 
$\sigma /\eta $ \par (m/s)& 
$\Theta $ \par (\r{ })& 
$v_{\rm i}^{\rm bulk} $ \par (m/s)& 
$v_{\rm ir}^{\rm bulk} $ \par & 
$v_{\rm ir}^{\rm conf} $ \par  \\
\hline
water& 
23& 
0.95& 
72.25& 
1.00& 
72.75& 
0& 
72.75& 
1.000& 
1.000 \\
\hline
n-C$_{\rm 16}$H$_{\rm 34}$& 
22& 
3.1& 
27.15 & 
0.77& 
8.77& 
5& 
8.77& 
0.121& 
0.128 \\
\hline
8OCB& 
85& 
9.6 & 
27.00 & 
0.98& 
2.81& 
5& 
2.80& 
0.039& 
0.036 \\
\hline
\end{tabular}
\caption{Bulk fluid parameters and imbibition speeds of the three 
liquids investigated. The notation is described in the text. (The bulk fluid 
parameters for water were taken from ref. \cite{Weast1981}, the ones of n-C$_{\rm 16}$H$_{\rm 34}$ from ref. \cite{Small1986} and the parameter of 8OCB from ref. \cite{Jadzyn2001, Langevin1976, Zywocinski1987}.)}
\label{tab1}
\end{center}
\end{table*}

\section{Summary}
We have demonstrated that simple capillary rise experiments in nanoporous 
hosts allow one, in principle, to study the relative magnitude of fluid 
parameters of liquids confined in nanopores. In the case of water, 
n-hexadecane and 8OCB we observe capillary rise velocities that scale as 
expected from the bulk fluid parameters. Thus, we find no hints of a 
breakdown of the continuous-like description of the hydrodynamic flow of 
these liquids in nanoporous silica. This result achieved in a capillary flow 
geometry, and thus in a flow experiment with Hagen-Poiseuille type flow 
geometry, is maybe not too surprising, if one recalls the observations 
reported for microscopic thin films between mica surfaces in the surface 
force apparatus setup (Couette flow geometry) \cite{Raviv2001,Israelachvili1988, Ruths2000}. There, deviations 
from the bulk fluidity have been reported for film thicknesses on the order 
of a couple of monolayers and below, only.\\ 

It is understood, that we can miss deviations in the absolute transport 
properties of the spatial confined liquids by the type of experiment and 
analysis presented. Our findings allow, however, only two alternative 
conclusions: Either the macroscopic description of the nanofluidity in Vycor 
is valid for all three liquids, or it is invalid, but deviates in the same 
direction and probably for a similar reason for all three substances. In 
fact, we assume that the second alternative is the correct one. Experiments 
on the absolute fluid transport properties of Vycor glass have been reported 
in the literature \cite{Debye1959,Abeles1990}, that indicate the existence of one or two strongly adsorbed immobile or less mobile layers of molecules at the silica pore walls. A finding which is reminiscent of the two distinct regimes that one finds in sorption isotherms of simple molecules, e.g. argon, in Vycor, i.e. an initial layer growth and a subsequent formation of pore condensate in the center of the pores \cite{Huber1999}. Presumably the microscopic boundary layer exists also for water, n-C$_{\rm 16}$H$_{\rm 34}$ and 8OCB in Vycor and decreases or increases the absolute flow rates through the pores depending on its mobility and the resulting slippage at the pore wall \cite{Urbakh2004, Drake1990, Granick2003, Gelb2002}.\\

Last but not least, we would like to mention, that the imbibition front in 
Vycor is a fine example of an advancing interface in a random environment \cite{Alava2004}. Therefore, imbibition experiments on the capillary rise of liquids in this porous host are also interesting in order to study kinetic roughening of such interfaces in nanostructured environments.\\

\begin{acknowledgements}
This work has been supported within the DFG priority program 1164
(Grant No. Hu 850/2).
\end{acknowledgements}


\end{document}